\documentclass{article}
\usepackage{amsmath}

\input{tcilatex}
\usepackage{graphicx}

\begin{document}

\title{Storage and perpendicular retrieving of two-dimensional pulses in
electromagnetically induced transparency media}
\author{Gor Nikoghosyan \\
Institute for Physical Research of\\
Armenian National Academy of Sciences\\
nikgor@ipr.sci.am}
\maketitle

\begin{abstract}
Propagation of two dimensional pulses in electromagnetically induced
tranparency media in the case of perpendicular storing and retrieving pulses
has been analyzed. It has been shown that propagation control of the pulses
in optically thick media can be used for producing interchange between pulse
time-shape and intensity profile distribution. A simple obvious analytical
solution for the retrieved new field has been obtained.
\end{abstract}

Electromagnetically induced transparancy (EIT) is a coherent interaction
process where a coupling laser field is used to modify the optical
properties of an atomic medium for probe laser field \cite{harris}. The
common EIT\ system is a medium of three level $\Lambda $\ atoms interacting
with two resonant laser fields tuned to two photon resonance. Since the
discovery of EIT\ great attention has been payed to that, which has been
caused by many unique applications and effects treated by this effect, such
as ultraslow light propagation \cite{hau}, light storage \cite{Fleisch},
coherent control of temporal pulse shaping \cite{italy} etc. By using EIT
also, the traditional effects of nonlinear optics may be enhanced \cite%
{stand}. The theoretical study presented in this paper discusses and
explains how to exploit the pecularities of EIT propagation dynamics for
both the determination of temporal shape of probe pulse and the
determination of probe pulse profile intensity distribution. Storaging and
retrieving pulses are assumed to propagate in perpendicular derections. The
possibility of changing the direction of the pulse by a swtich of control
beam direction was already shown in \cite{swed}. It will be shown below that
the retrieved pulse time-shape and profile intensity distribution contains
information about the initial probe pulse profile and time-shape
respectively. Similar results has been obtained in atomic beams \cite%
{masalas}.

A medium of three level $\Lambda $\ atoms with two metastable lower states
is considered (Fig.1). State $|2\rangle $ connects to state $|3\rangle $ by
coupling fields $E_{c1}=A_{c1}\cos \left( k_{c1}x-\omega _{c1}t+\varphi
_{c1}\right) $, $E_{c2}=A_{c2}\cos \left( k_{c2}y-\omega _{c2}t+\varphi
_{c2}\right) $, and state $|1\rangle $ connects to state $|3\rangle $ by
initial probe field $E_{p}=A_{p}\cos \left( k_{p}x-\omega _{p}t+\varphi
_{p}\right) $, and the generated new field $E_{n}=A_{n}\cos \left(
k_{p}y-\omega _{p}t+\varphi _{n}\right) $ which appears only after $E_{c2}$
turns on. The field $E_{c2}$ assumed turn on at $t=\tau _{1}$ when $E_{c1}$
has been turned off already. Fields $E_{c1}$ and $E_{c2}$ are assumed to be
polarized along z axis and wave vectors of probe and the storing field are
assumed to be approximately equal $k_{p}\approx k_{c1}$ \cite{swed}. So the
system Hamiltonian in rotating wave aproximation for $t<\tau _{1}$ when only
probe $E_{p}$ and storing field $E_{c1}$ are turned on is as follows:

\begin{equation}
H=\hbar \sigma _{33}\Delta -\hbar \Omega _{p}\sigma _{31}-\hbar \Omega
_{c1}^{\ast }\sigma _{32}+H.c  \label{hamil}
\end{equation}

Where $\Omega _{i}=\dfrac{A_{i}\mu _{3i}}{\hbar }$ is Raby frequency.

With th Hamiltonian (\ref{hamil}) the atom dynamics is described by the
Bloch equation:

\begin{equation}
\overset{.}{\rho }=-\frac{i}{\hbar }\left[ \rho H\right] +\Lambda \rho
\label{bloch}
\end{equation}

Where $\Lambda $ is dissipation matrix, which discribes decoherence and
spontaneous emission.\ Under assumption

\begin{equation}
\Omega _{p}\ll \Omega _{c1}\text{ and }\Omega _{i}T\gg 1\text{,}
\label{aprox}
\end{equation}

which means that the probe pulse is weak compared to storing pulse, and the
interaction is supposed to be adiabatic.

The propagation equation for the Rabi frequencies $\Omega _{p}$and $\Omega
_{c1}$ may be written as

\begin{eqnarray}
\left( \frac{\partial }{\partial x}+\frac{1}{c}\frac{\partial }{\partial t}%
\right) \Omega _{p} &=&iq_{p}\rho _{31}\text{,}  \label{maxwel} \\
\left( \frac{\partial }{\partial x}+\frac{1}{c}\frac{\partial }{\partial t}%
\right) \Omega _{c1} &=&iq_{c}\rho _{32}\text{,}  \notag
\end{eqnarray}

where $q_{i}=\dfrac{2\pi \omega _{i}\mu _{i}^{2}N}{\hbar c}$.

To the lowest nonvanishing order with respect to (\ref{aprox}) the Bloch
equation (\ref{bloch}) reduces to

\begin{eqnarray}
\rho _{21} &=&-\frac{\Omega _{p}}{\Omega _{c1}},\ \rho _{31}=-\frac{i}{%
\Omega _{c1}}\frac{\partial }{\partial t}\rho _{21}\text{,}  \label{reduc} \\
\rho _{11} &=&1,\ \rho _{22}=\rho _{33}=\rho _{32}=0\text{,}  \notag
\end{eqnarray}

Where assumption that all atoms of the medium are initially in the state 1
is made.

By substituiting (\ref{reduc}) into the Maxwell equation (\ref{maxwel}) \
one gets pulses propagation equations for the pulses:

\begin{eqnarray}
\left( \frac{\partial }{\partial x}+\frac{1}{c}\frac{\partial }{\partial t}%
\right) \Omega _{p} &=&-\frac{q_{p}}{\Omega _{c1}}\frac{\partial }{\partial t%
}\frac{\Omega _{p}}{\Omega _{c1}}  \label{propa} \\
\left( \frac{\partial }{\partial x}+\frac{1}{c}\frac{\partial }{\partial t}%
\right) \Omega _{c1} &=&0  \notag
\end{eqnarray}

Solution of (\ref{propa}) can be found easily:

\begin{equation}
\Omega _{p}\left( t,x,y\right) =\Omega _{c}\left( t-x/c\right) f\left(
x-\int\limits_{0}^{t-x/c}\frac{\Omega _{c}^{2}\left( t^{\prime }\right) }{%
q_{p}}dt^{\prime }\right) a\left( y\right) \text{,}  \label{dsp}
\end{equation}

\begin{equation}
\Omega _{c}\left( t,x\right) =\Omega _{c}\left( t-x/c\right) \text{,}
\label{coupl}
\end{equation}

In (\ref{dsp}) f(x) is determined by probe pulse initial time shape, a(y) by
probe pulse profile intensity distribution. It was also assumed coupling
pulse profile is wide so its profile intensity distribution can be neglected.

Propagation of coherence $\rho _{21}$ is found by substituiting equations (%
\ref{dsp}) \ and (\ref{coupl}) into (\ref{reduc}):

\begin{equation}
\rho _{21}=-f\left( x-\int\limits_{0}^{t-x/c}\frac{\Omega _{c}^{2}\left(
t^{\prime }\right) }{q_{p}}dt^{\prime }\right) a\left( y\right) \text{.}
\label{coh}
\end{equation}

Equations (\ref{dsp}), (\ref{coupl}) describes the well-known propagation of
probe pulse with nonlinear group velocity $q_{p}/\Omega _{c}^{2}$. In
particular, if we turn off coupling pulse, probe's group veloctity tends to
zero, but if we turn on coupling pulse again (retrieving pulse), the probe
pulse can be obtained since the coherence $\rho _{21}$ stores information
about the initial probe.

Lets now consider the case when the retrieving coupling is turned on
perpendicularly to the initial pulse propagation direction. It is supposed
that retrieving coupling pulse $\Omega _{c2}$ turns on after turning off
storing pulse $\Omega _{c1}$. $\Omega _{c2}\neq 0$ for $t>\tau _{1}$. The
system Hamiltonian for $t>\tau _{1}$ is

\begin{equation}
H=\hbar \sigma _{33}\Delta -\hbar \Omega _{n}\sigma _{31}-\hbar \Omega
_{c2}\sigma _{32}+H.c  \label{newham}
\end{equation}

It is seen that the new Hamiltonian (\ref{newham}) of the system is similar
to the Hamiltonian (\ref{hamil}), so the system dynamics is governed by the
same solution, but now we have different initial conditions. There is no new
field $\Omega _{n}$ at the medium entrance but\ because of the light storage
process medium has a certain coherence $\rho _{21}$. The new field and
coherence propagation equations can be obtained by the same procedure as
above.

\begin{equation}
\Omega _{n}=\Omega _{c2}\left( t-y/c\right) g\left( y-\int\limits_{\tau
_{1}}^{t-y/c}\frac{\Omega _{c2}^{2}\left( t^{\prime }\right) }{q_{p}}%
dt^{\prime }\right) b(x)  \label{newfield}
\end{equation}

\begin{equation}
\rho _{21}=-g\left( y-\int\limits_{\tau }^{t-y/c}\frac{\Omega
_{c2}^{2}\left( t^{\prime }\right) }{q_{p}}dt^{\prime }\right) b(x)
\label{newcoh}
\end{equation}

Where $g$ describes the temporal shape of the generated new field and $%
b\left( x\right) $ describes the intensity profile distribution of the
generated field, it has been supposed again that the coupling pulse is wide
enough to neglect its profile distribution. Note, that

\begin{equation}
\int\limits_{\tau _{1}}^{t-y/c}\frac{\Omega _{c2}^{2}\left( t^{\prime
}\right) }{q_{p}}dt^{\prime }=0\text{ for }t<\tau _{1}  \label{condit1}
\end{equation}

since $\Omega _{c2}$ turns on later than $\tau _{1}$, and

\begin{equation}
\int\limits_{0}^{t-x/c}\frac{\Omega _{c1}^{2}\left( t^{\prime }\right) }{%
q_{p}}dt^{\prime }=\int\limits_{0}^{\tau _{1}-x/c}\frac{\Omega
_{c1}^{2}\left( t^{\prime }\right) }{q_{p}}dt^{\prime }fort\geq \tau _{1}
\label{condit2}
\end{equation}

since $\Omega _{c1}$ turns off to the moment $\tau _{1}.$

\qquad Now, by comparing (\ref{newcoh}) and (\ref{coh}) at the moment $%
t=\tau _{1}$ and taking into acount (\ref{condit1}) and (\ref{condit2}), one
obtains for the generated new field time shape and intensity profile
distribution:

\begin{eqnarray*}
g\left( y\right) &=&a\left( y\right) \\
b\left( x\right) &=&f\left( x-\int\limits_{0}^{\tau _{1}-x/c}\frac{\Omega
_{c1}^{2}\left( t^{\prime }\right) }{q_{p}}dt^{\prime }\right)
\end{eqnarray*}

This means that the obtained new field has interchanged time shape and
intensity profile distribution with the initial probe field. Or, in other
words, temporal shape of the generated new field is governed by the initial
probe pulse intensity profile distribution and the intensity profile
distribution of the generated new pulse is governed by the initial pulse
temporal shape. Finally intrinsic a simple obvious solution can be obtained
for the new pulse propagation.

\begin{equation}
\Omega _{n}=\Omega _{c2}\left( t-y/c\right) a\left( y-\int\limits_{\tau
_{1}}^{t-y/c}\frac{\Omega _{c2}^{2}\left( t^{\prime }\right) }{q_{p}}%
dt^{\prime }\right) f\left( x-\int\limits_{0}^{\tau _{1}-x/c}\frac{\Omega
_{c1}^{2}\left( t^{\prime }\right) }{q_{p}}dt\right)  \label{newpulse}
\end{equation}

Where $\Omega _{c2}\left( t-y/c\right) a\left( y-\int\limits_{\tau
_{1}}^{t-y/c}\dfrac{\Omega _{c2}^{2}\left( t^{\prime }\right) }{q_{p}}%
dt^{\prime }\right) $ describes the temporal shape of the new field where $%
a\left( y\right) $ is initial profile distribution of the probe and $f\left(
x-\int\limits_{0}^{\tau _{1}-x/c}\dfrac{\Omega _{c1}^{2}\left( t^{\prime
}\right) }{q_{p}}dt\right) $ describes the profile intensity distribution of
the new field where $f\left( t\right) $ is the time-shape of the initial
probe. Equation (\ref{newpulse}) shows that mesuring generated field
time-shape gives us information about the initial probe pulse profile and
vice versa mesuring of generated pulse profile gives us information about
the initial probe pulse time-shape:

Below it will be introduced the probe field, new field and coherence spatial
distributions at the different time moments. Probe pulse temporal shape and
profile are taken as $f\left( t\right) \Omega _{c}\left( t\right) T=0.9e^{-(%
\frac{t}{T}-2.5)^{2}}+1.2e^{-(\frac{t}{T}-6)^{2}}$, $a\left( y\right) =e^{-(%
\frac{y}{20x_{0}}-5)^{2}}$ respectively (Fig.2 a,b). Time shape of storing
pulse and retrieving pulses were assumed as $\Omega
_{c1}T=\{t>8,10e^{-\left( \frac{t}{T}-8\right) ^{2}},t<8,10\}$, $\Omega
_{c2}T=\{t<15,10e^{-\left( \frac{t}{T}-8\right) ^{2}},t>15,10\}$ (Fig.2c),
where T is the duration of probe pulse and $x_{0}=\left( qT\right) ^{-1}$.
The spatial distribution of the probe pulse is depicted on Figs 3.1a-3.5a.
As can be seen, probe pulse propagates along x axis and is absorbed when
cupling pulse is turned off. But due to the coherence pumped to the media
(Figs 3b,d) the new pulse can be generated by switching on the coupling
(Figs 3c) . Since coherence is distributed along both x and y axis, even
when retrieving pulse turns on along y axis new pulse is obtained. As
stated, above the generated new pulse propogation direction coincides with
retrieving pulse propogation direction and the time-shape is governed by the
initial probe pulse profile while the profile of the new pulse is governed
by the time-shape of the initial probe pulse.

For 1cm long $Rb^{85}$ cell at an atom density $10^{14}atoms/cm^{3}$ if
probe and coupling are applied at D1 hyperfine transitions $x_{0}\approx
0.002cm$ for $T=10^{-6}\sec $. One can see from figures that proposed
experiement can be implemented with laser width of about mm.

In conclusion it has been shown that propagation control of the pulses in
optically thick media can be used for producing interchange between pulse
time-shape and intensity profile distribution. This mechanism is based on
light storage and retrieving process.in the case of perpendicular
propagation directions of storing and retreiving pulses. This effect can
find application both in mesuring pulse time-shape and for mesuring pulse
profile. Analytical solution for the generated new field has been obtained
and numerical results describing the dynamics of the system in detail has
been presented.

Author gratefully acknowledges G.Grigoryan for useful discussions.

\clearpage

\begin{center}
{\huge Figures}

FIG.1 A medium of three level $\Lambda $\ atoms with two metastable lower
states. State $|2\rangle $ connects to state $|3\rangle $ by coupling fields 
$E_{c1}=A_{c1}\cos \left( k_{c1}x-\omega _{c1}t+\varphi _{c1}\right) $, $%
E_{c2}=A_{c2}\cos \left( k_{c2}y-\omega _{c2}t+\varphi _{c2}\right) $, and
state $|1\rangle $ connects to state $|3\rangle $ by initial probe field $%
E_{p}=A_{p}\cos \left( k_{p}x-\omega _{p}t+\varphi _{p}\right) $, and the
generated new field $E_{n}=A_{n}\cos \left( k_{p}y-\omega _{p}t+\varphi
_{n}\right) $ which appears only after $E_{c2}$ turns on and propagates in
the same direction.

\includegraphics[height=3in]{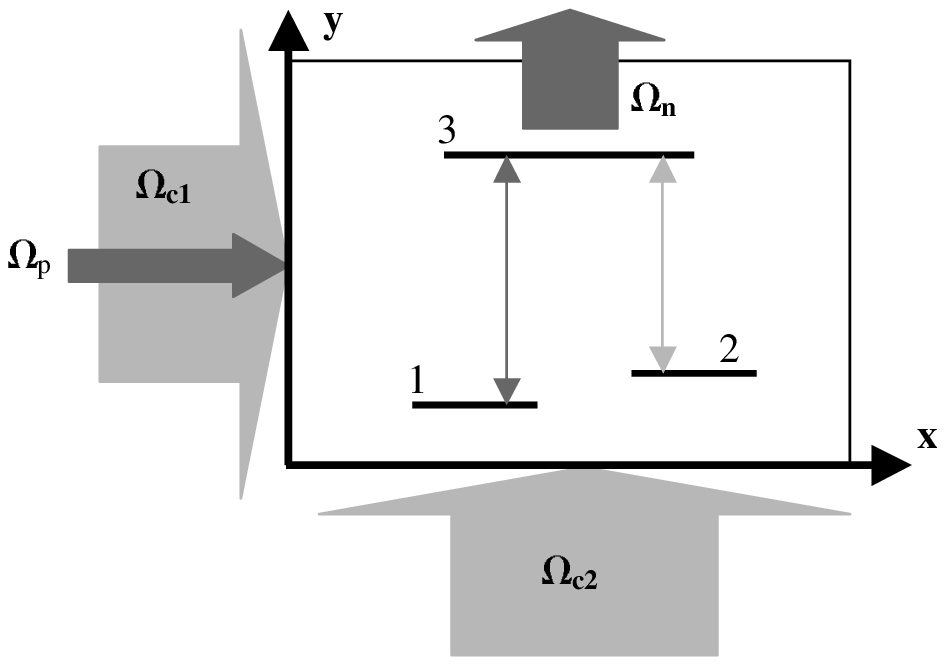}

\clearpage

FIG.2 Probe pulse temporal shape and profile $f\left( t\right) \Omega
_{c}\left( t\right) T=0.9e^{-(\frac{t}{T}-2.5)^{2}}+1.2e^{-(\frac{t}{T}%
-6)^{2}}$, $a\left( y\right) =e^{-(\frac{y}{20x_{0}}-5)^{2}}$ (a) (b)
respectively. Time shapes of storing and retrieving pulses (c) $\Omega
_{c1}T=\{t>8,10e^{-\left( \frac{t}{T}-8\right) ^{2}},t<8,10\}$, (solid) $%
\Omega _{c2}T=\{t<15,10e^{-\left( \frac{t}{T}-8\right) ^{2}},t>15,10\}$
(dashed).

\includegraphics[height=1.5in]{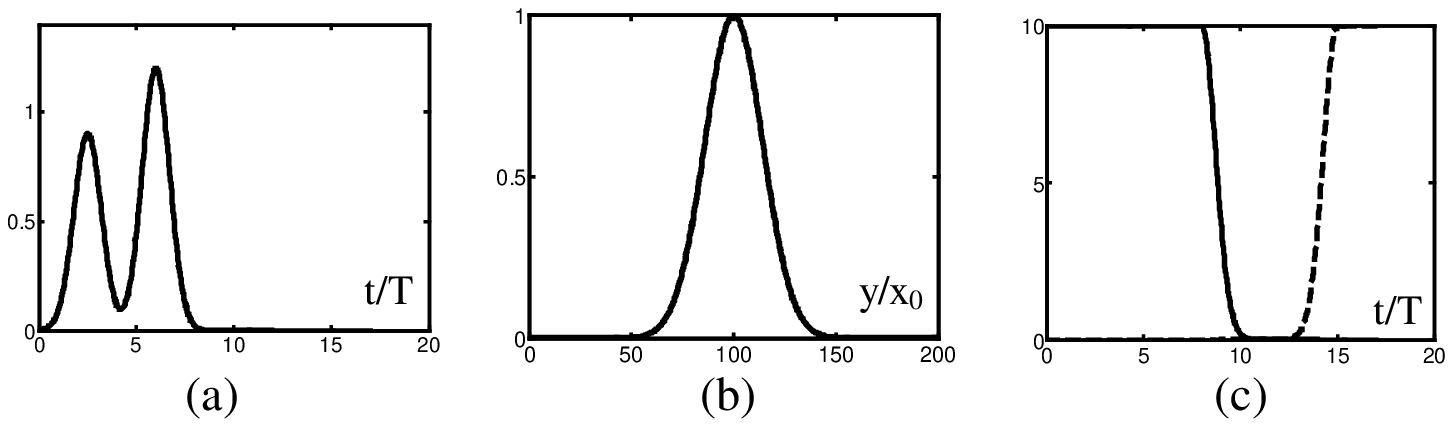}

\clearpage

FIG.3 Probe field (a), new field (c) and coherence (b,d) spatial
distributions at the different moments of time.

\includegraphics[height=7in]{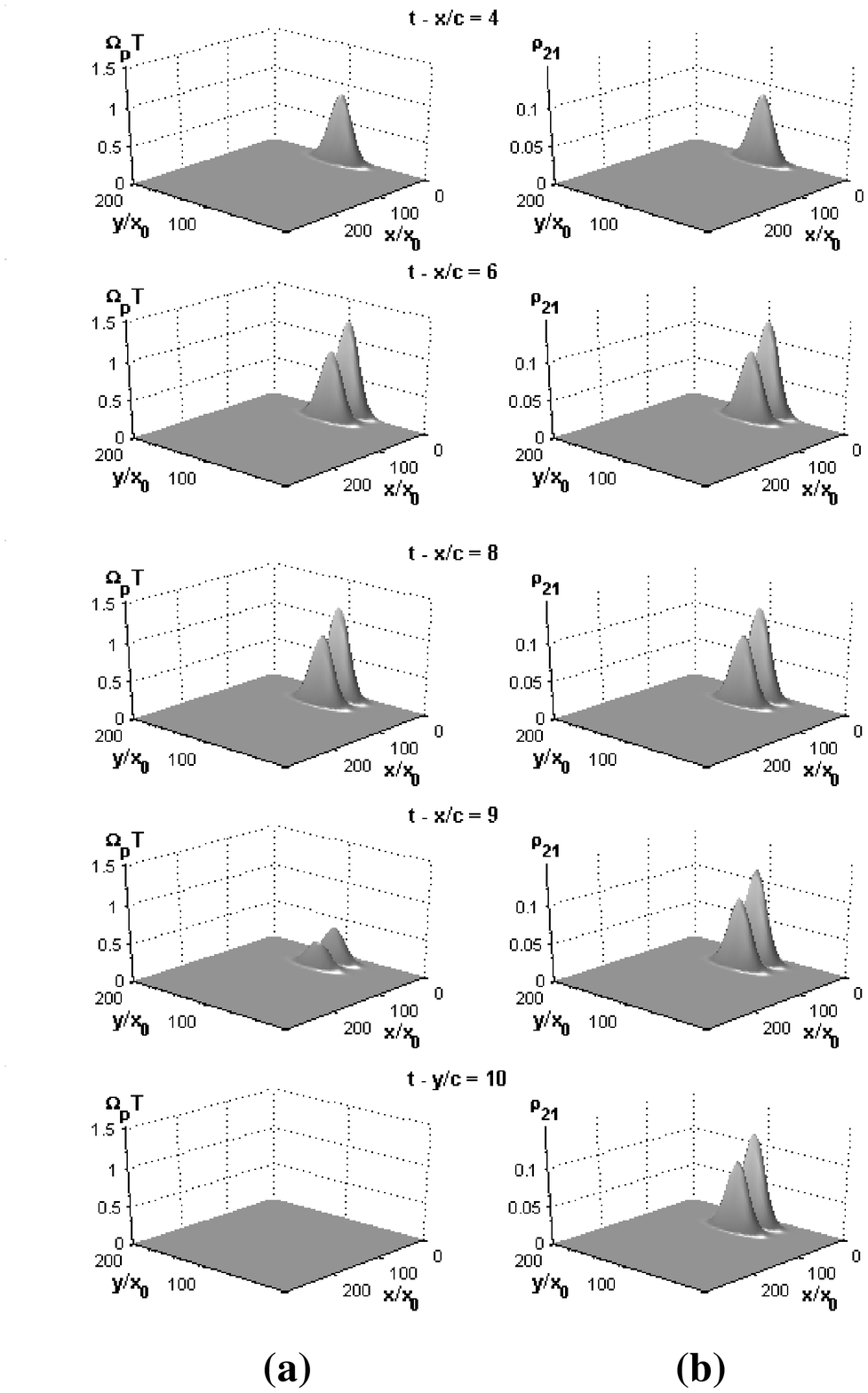}
\includegraphics[height=7in]{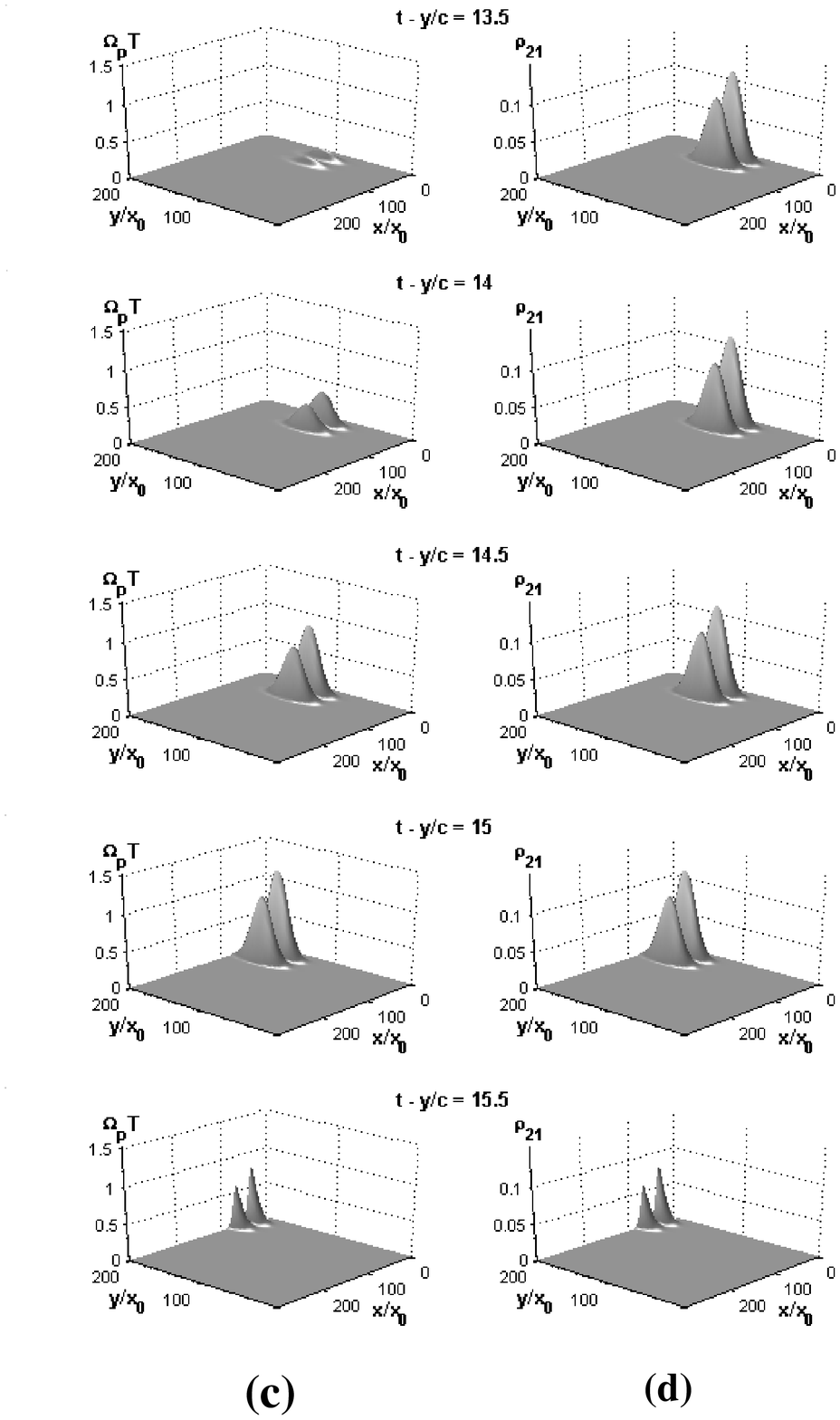}

\end{center}
\end{document}